\documentclass[aps,prb,twocolumn,english,superscriptaddress,showpacs,floatfix,groupedaddress]{revtex4}
\usepackage{amsmath}
\usepackage{amssymb}
\usepackage{epsfig,graphics,graphicx}
\usepackage{babel}

\begin{document}

\title{Mesoscopic Multi-Particle Collision Dynamics of Reaction-Diffusion Fronts}

\author{Kay Tucci}
\email[]{kay@ula.ve}
\affiliation{SUMA-CeSiMo, Universidad de Los Andes,
M\'erida 5101, Venezuela}
\author{Raymond Kapral}
\email[]{rkapral@chem.utoronto.ca}
\affiliation{Chemical Physics Theory Group, Department of
Chemistry, University of Toronto, Toronto, ON M5S 3H6, Canada}

\date{\today}

\begin{abstract}
A mesoscopic multi-particle collision model for fluid dynamics is
generalized to incorporate the chemical reactions among species
that may diffuse at different rates. This generalization provides
a means to simulate reaction-diffusion dynamics of complex
reactive systems. The method is illustrated by a study of cubic
autocatalytic fronts. The mesoscopic scheme is able to reproduce
the results of reaction-diffusion descriptions under conditions
where the mean field equations are valid. The model is also able
to incorporate the effects of molecular fluctuations on the
reactive dynamics.
\end{abstract}

\pacs{02.70.Ns, 05.10.Gg, 05.20.Dd}

\maketitle

\section{Introduction}
Mesoscopic models provide coarse-grained descriptions of the
dynamics of systems that neglect certain details at microscopic
scales while retaining essential dynamical features at mesoscopic
and macroscopic scales. Consequently, a convenient way to study of
the dynamics of complex systems over a large range of interesting
space and time scales is through the use of such models. In
physical and biological systems we often encounter situations
where mean field descriptions of reactions break down and
molecular fluctuations play an important role in determining the
character of the system's dynamics. Such effects are especially
relevant for reactions taking place in nano-scale domains or
biochemical reactions at the cellular level. Fluctuations also
play a role in far-from-equilibrium systems near bifurcation
points or when the system behaves chaotically since the system is
especially susceptible to perturbations in such regimes.
\cite{nicolis}  Mesoscopic models are able to capture the
influence of such molecular fluctuations on the dynamics.
Mesoscopic models are also useful for simulating the dynamics of
macroscopic systems because they often provide stable
particle-based simulation schemes and can be implemented in
complex geometries.

In this article we consider a generalization of a mesoscopic
multi-particle collision (MPC) (or stochastic rotation) model
\cite{mpc1,mpc2,mpc3} to a pattern-forming chemically reacting
system. We show how the multi-particle collision rule can be
generalized to a multi-component system to yield different
diffusion coefficients for the chemical species. Differences in
diffusion coefficients can give rise to chemical instabilities
which cannot occur if the diffusion coefficients of all species
are equal. Reactions are incorporated, also at a mesoscopic level,
by combining a birth-death description of reactive events with
multi-particle collisions. The mesoscopic dynamics preserves all
the basic conservation laws of the system and leads to the
macroscopic evolution laws on long distance and time scales.

To illustrate the scheme, the reactive MPC dynamics is used to
investigate the evolution and structure of a cubic autocatalytic
front. The cubic autoatalytic reaction is $A+2B \to 3B$, where the
autocatalyst $B$ consumes the fuel $A$. If one considers a
two-dimensional rectangular domain (or a thin rectangular slab in
three dimensions) with $B$ in left portion and $A$ in the right
portion, a reaction front will propagate from left to right. While
the simulations presented in this paper are for cubic
autocatalytic fronts, the manner in which the diffusion process is
modelled to yield different diffusion coefficients for different
chemical species and the way reactions are incorporated in the
model presage extensions of the theory and applications to more
complex far-from-equilibrium reactive systems.

The paper is organized as follows: In Sec.~\ref{sec:mesomodel} we
sketch the basic elements of the multi-particle collision model
and present its generalization to reactive systems where the
chemical species can have different diffusion coefficients.
Section~\ref{sec:front} describes the simulation of cubic
autocatalytic fronts and compares the results of the mesoscopic
simulations with the predictions of reaction-diffusion equations.
The conclusions of the paper are given in Sec.~\ref{sec:conc}.

\section{Mesoscopic Model} \label{sec:mesomodel}

In multi-particle collision dynamics a system containing $N$
particles with continuous positions ${\bf r}_i$ and velocities
${\bf v}_i$ evolves through a sequence of free streaming and
collision steps~\cite{mpc2}. The collisions among the particles
take place in the following way: the system is divided into cells
and at time intervals $\tau$ each cell labelled by $\xi$ is
assigned at random a rotation operator $\hat{\omega}_\xi$ from
some suitable set of rotation operators. The center of mass
velocity ${\bf V}_\xi$ of the particles in cell $\xi$ is computed
and the post-collision velocity ${\bf v}'_i$ of particle $i$ in
the cell is determined by rotating its velocity, relative to the
cell center of mass velocity, and adding the center of mass
velocity to the result of this rotation:
\begin{equation}
{\bf v}'_i= {\bf V}_\xi +\hat{\omega}_\xi({\bf v}_i-{\bf V}_\xi)\;.
\label{eq:mpc}
\end{equation}
The velocity of every particle in cell $\xi$ is rotated by the
same rotation operator but the rotation operator varies from cell
to cell. The dynamics then consists free streaming interspersed by
these multi-particle collision events. It has been shown that this
dynamics conserves mass, momentum and energy and thus leads to the
full set of Navier-Stokes equations on long distance and time
scales~\cite{mpc2,mpc3,mesofin}. The method has been applied to the study
of a variety of systems \cite{mesofin} including hydrodynamic
flows \cite{flows}, colloids \cite{colloids}, polymers
\cite{polymers}, Brownian motion \cite{songhi} and simple
diffusion-influenced reaction dynamics \cite{kay}.

We present a generalization of this model that allows the dynamics
of reaction-diffusion systems to be investigated. This
generalization entails several extensions of the MPC model. In
particular, a multi-component version of the MPC model
\cite{kay,yeomans} must be constructed that accounts for reactions
among the chemical species and allows for the possibility that the
diffusion coefficients of the species differ.

\subsection*{Diffusion}

A multi-component MPC dynamics that provides a simple way to
control the diffusion coefficients of different chemical species
can be constructed as follows. Suppose we have $s$ species
labelled by an index $\alpha$. Instead of applying the MPC
operator to all particles in a cell, we assume that multi-particle
collision operators act to change the velocities of a fraction of
the particles of species $\alpha$ in a cell for
$\alpha=1,\dots,s$. More specifically, in each cell $\xi$ each
particle of species $\alpha$ is chosen with probability
$\gamma_\alpha$. If ${\bf v}_i^\alpha$ is the velocity of a chosen
particle $i$ of species $\alpha$ and ${\bf V}_\xi^c$ is the center
of mass velocity of all chosen particles, the post-collision
velocities of those particles that undergo collision are given by
\begin{equation} {\bf v}^{\alpha \prime}_i= {\bf V}_\xi^c
+\hat{\omega}_\xi({\bf v}_i^\alpha-{\bf V}_\xi^c)\;.
\label{eq:mpc2}
\end{equation}
The post-collision velocities of the particles that do not take
part in the multi-particle collision do not change.  The diffusion
coefficients $D_\alpha$ are functions of
$\{\gamma_{\alpha'}|\alpha'=1,\dots,s\}$, which can be tuned to
change the values of the diffusion coefficients.

\begin{figure}[htbp]
\centerline{\mbox{
\epsfig{file=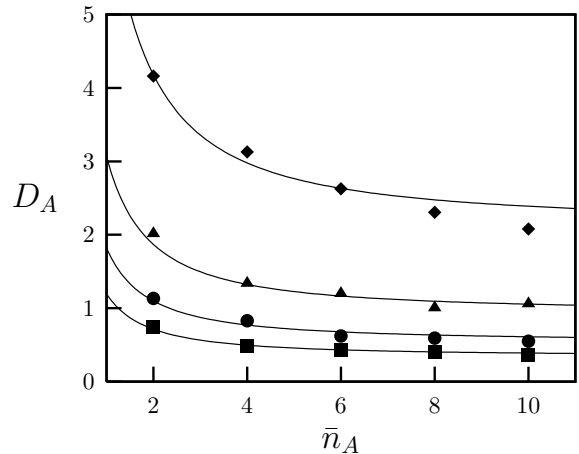,width=0.75\linewidth,clip=,angle=90}
}} \caption{Diffusion coefficient $D_A(\gamma_A)$ as a function of
the density $\bar{n}_A$ for various values of $\gamma_A$. Squares,
$\gamma_A = 1.00$. Circles, $\gamma_A = 0.75$. Triangles, $\gamma_A
= 0.50$. Diamonds $\gamma_A = 0.25$. Solid lines plot the
theoretical value (\protect Eq.~(\ref{eq:diffusionCoeficient})) of
the diffusion coefficient.
\label{fig:D-vs-na} }
\end{figure}

In order to investigate the range over which the diffusion
coefficients can vary, we consider the self diffusion coefficient
of a single species $A$ and change both the mean particle density
$\bar{n}_A$ and the fraction $\gamma_A$ of particles that
participate in the multi-particle collisions.
Figure~\ref{fig:D-vs-na} plots $D_A(\gamma_A)$, determined from
the slope of the mean square displacement versus time, as a
function density $\bar{n}_A$ for different values of $\gamma_A$.
>From these results one sees that the self diffusion coefficient
can be varied by about a factor of five by changing the values of
$\gamma_A$ at a fixed density.

The self diffusion coefficient for $\gamma_A \ne 1$ can be
estimated in the Boltzmann approximation where correlations are
neglected. The discrete-time Green-Kubo expression for the
diffusion coefficient is \cite{mpc3,kay}
\begin{equation}
D_A(\gamma_A)= \frac{1}{2}\langle v_x^2 \rangle +\sum_{n=1}^\infty
\langle v_x(n) v_x \rangle \;,
\end{equation}
where, without loss of generality, we have set $\tau=1$.  Taking
into account the collision rule where, on average, a fraction
$\gamma_A$ of the particles undergo multi-particle collisions and
fraction $1-\gamma_A$ do not, we have
\begin{equation}
D_A(\gamma_A)= \frac{1}{2}\langle v_x^2 \rangle
+(1-\gamma_A)\langle v_x^2 \rangle +\gamma_A \langle v_x^{(1)}v_x
\rangle +\dots \;,
\end{equation}
where $v^{(1)}_x$ is the post-collision value of the velocity at
time $\tau=1$. Assuming that higher order collision terms can be
expressed in terms of the first collision so the series is
geometric, we obtain
\begin{eqnarray}
D_A(\gamma_A)&=&-\frac{1}{2}\langle v_x^2 \rangle +\langle v_x^2
\rangle \Big( 1+(1-\gamma_A) +\gamma_A r_D +\dots\Big) \nonumber
\\
&\approx& -\frac{1}{2}\langle v_x^2 \rangle +\frac{\langle v_x^2
\rangle }{\gamma_A (1-r_D)} \nonumber \\
&=& \frac{\langle v_x^2 \rangle}{2}
\left[\frac{2-\gamma_A(1-r_D)}{\gamma_A(1-r_D)}\right]\;,
\label{eq:diffusionCoeficient}
\end{eqnarray}
where
\begin{equation}
r_D=\frac{\langle v_x^{(1)}v_x \rangle}{\langle v_x^2 \rangle}=
\frac{2(1-e^{-\bar{n}_A})+\bar{n}_A}{3\bar{n}_A}\;,
\end{equation}
was computed in Ref.~\cite{kay}. The comparison in
Fig.~\ref{fig:D-vs-na} shows that this analytical expression
(solid lines) accurately describes the simulation data.

\subsection*{Reaction}

The mesoscopic dynamics must also be generalized to allow for
chemical reactions among the species. Our earlier study of
diffusion-influenced reactions \cite{kay} was restricted to a
simple $A +C \rightleftharpoons B +C$ reaction that occurs when
the $A$ or $B$ particles collide with catalytic spheres $C$. Since
we are now interested in reactions that occur among the mesoscopic
particles, we instead use a birth-death stochastic law to describe
the reactive events~\cite{nicolis,gardiner}.

Here we restrict our considerations to the cubic autocatalytic
reaction $A+2B \rightarrow 3B$. Independently in each cell we
assume the reaction takes place with probability $p_R = k n_{A}
n_{B} (n_{B} - 1)$, where $n_{\alpha}$ is the number of molecules
of species $\alpha$ in a cell. The reactive dynamics in a cell is
described by the Markov chain, \cite{ebeling}
\begin{equation}
P({\bf n},t+1)=\sum_{{\bf n}'}W({\bf n}|{\bf n}') P({\bf n}',t)\;,
\end{equation}
where ${\bf n}=(n_A,n_B)$, $P({\bf n},t)$ is the probability that
there are ${\bf n}$ particles in the cell at time $t$ and the
transition matrix $W$ is given by
\begin{eqnarray}
&&W({\bf n}|{\bf n}')= kn_A'n_B'(n_B'-1) \delta_{n_A, n_A'-1}
\delta_{n_B, n_B'+1} \nonumber \\
&& \qquad +(1- kn_A'n_B'(n_B'-1))\delta_{n_A, n_A'} \delta_{n_B, n_B'}\;.
\end{eqnarray}
The (discrete time) rate of change of the mean density of species
$\alpha$, $\bar{n}_\alpha(t)= \sum_{{\bf n}}n_\alpha P({\bf
n},t)$, is
\begin{eqnarray}
\bar{n}_\alpha(t+1)-\bar{n}_\alpha(t)= \sum_{{\bf n},{\bf n}'}
n_\alpha \Big(W({\bf n}|{\bf n}') - \delta_{{\bf n},{\bf n}'}\Big)
P({\bf n},t). \label{eq:meanfield}
\end{eqnarray}
We assume that the MPC non-reactive collisions are sufficiently
effective to maintain a local equilibrium Poissonian distribution in the
cells so that
\begin{equation}
P({\bf n},t) \approx P_\ell(n_A;\bar{n}_A(t)) P_\ell(n_B;\bar{n}_B(t))\;,
\end{equation}
where the local Poisson distribution is $P_\ell(n;\bar{n}(t))=e^{-\bar{n}(t)}
\bar{n}(t)^{n}/n!$. If we insert the local Poissonian approximation for
$P({\bf n},t)$ in the right hand side of Eq.~(\ref{eq:meanfield}) for
$\alpha=A$ we
obtain the discrete-time mean-field rate law,
\begin{eqnarray}
\bar{n}_A(t+1)-\bar{n}_A(t)= -k \bar{n}_A(t)\bar{n}_B^2(t) \;.
\label{eq:ratelaw}
\end{eqnarray}
A similar equation can be derived for species $B$. Thus, the mass
action rate law will describe the dynamics provided diffusion is
sufficiently rapid compared to reaction so that a local Poissonian
distribution of particles is maintained during the evolution of
the reactive system. In this limit the discrete-time rate law will
closely approximate the continuous-time mass action rate law.

After the reaction step, the particles free stream using the
post-collision values of the velocities, taking into account the
boundary conditions of the system. Once all the particles have
been moved, the time advances one unit and the multi-particle
collision and reaction steps are applied again. This mesoscopic
dynamics conserves the total mass, momentum and energy of the system.

\section{Simulation of Chemical Fronts} \label{sec:sim}
In this section we show that the mesoscopic MPC model can be used
simulate the dynamics of cubic autocatalytic fronts on macroscopic
scales where comparisons with the predictions of
reaction-diffusion equations can be made. Cubic autocatalytic
fronts have been studied often in the context of a coupled pair of
reaction-diffusion equations for the $A$ and $B$ species.
\cite{cub1,cub2,cub3,cub4,cub5,cub6,mck} The particular focus of
many of these studies was on the transverse front instability that
occurs when the diffusion coefficient of the fuel is sufficiently
larger than that of the autocatalyst: at a critical value of the
diffusion coefficient ratio an instability will develop and the
planar front will become nonplanar and exhibit complex dynamics.

Our investigations will be confined to a simpler case of a binary
mixture undergoing the cubic autocatalytic reaction. For such a
reacting mixture the relevant macroscopic field variables are the
total mass density $\rho({\bf r},t)=\rho_A+\rho_B$, the local
concentration $c({\bf r},t)=\rho_A/\rho$, the center of mass
velocity ${\bf v}({\bf r},t)$ and the energy density $e({\bf
r},t)$. For the isothermal cubic autocatalytic reaction with no
net fluid flow so that ${\bf v}({\bf r},t)=0$, and taking equal
masses for the $A$ and $B$ species,  the macroscopic equation for
the number density of $A$ is \cite{deGM},
\begin{eqnarray}
\frac{\partial }{\partial t}\bar{n}_A({\bf r},t)= -k \bar{n}_A
\bar{n}_B^2 +D \nabla^2 \bar{n}_A\;, \label{eq:rd}
\end{eqnarray}
where $D$ is the mutual diffusion coefficient. The equation for
$\bar{n}_B({\bf r},t)$ is not independent and follows from number
conservation, $\bar{n}_A+\bar{n}_B=\bar{n}_0$.

\subsection*{Front profile} \label{sec:front}

The simulations of the reaction front using the MPC model were
carried out in a rectangular prism with length $l=200$ along $x$,
width $w=200$ along $y$ and height $h=5$ units along $z$. The
system was open along its length $x$, periodic boundary conditions
were imposed in the $y$-direction, and bounce-back reflection
boundary conditions were imposed on the top and bottom of the
prism along $z$. In order to initiate a chemical front, $A$
particles were distributed uniformly in the right side of the
prism, ($\bar{n}_A(x\ge 100)=10$, $\bar{n}_B(x \ge 100)=0$), while
$B$ particles were uniformly distributed in left side of the prism
($\bar{n}_A(x<100)=0$, $\bar{n}_B(x<100)=10$). The velocities were
chosen from a Maxwell-Boltzmann distribution with reduced temperature
$k_BT=\beta^{-1}=1/3$.

\begin{figure}[!htbp]
\centerline{\mbox{ \hfill \epsfig{file=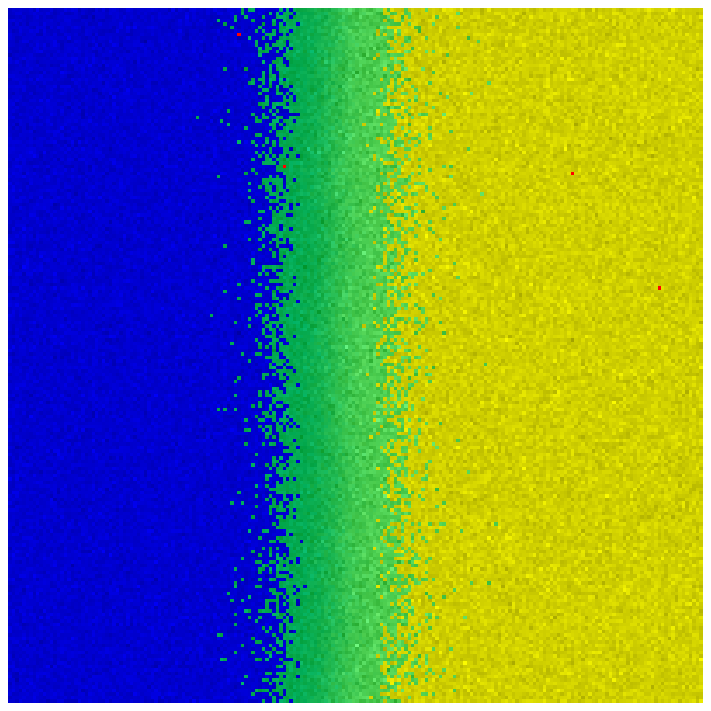,
        width=0.4\linewidth,clip=,angle=0}
\hspace{.05cm} \epsfig{file=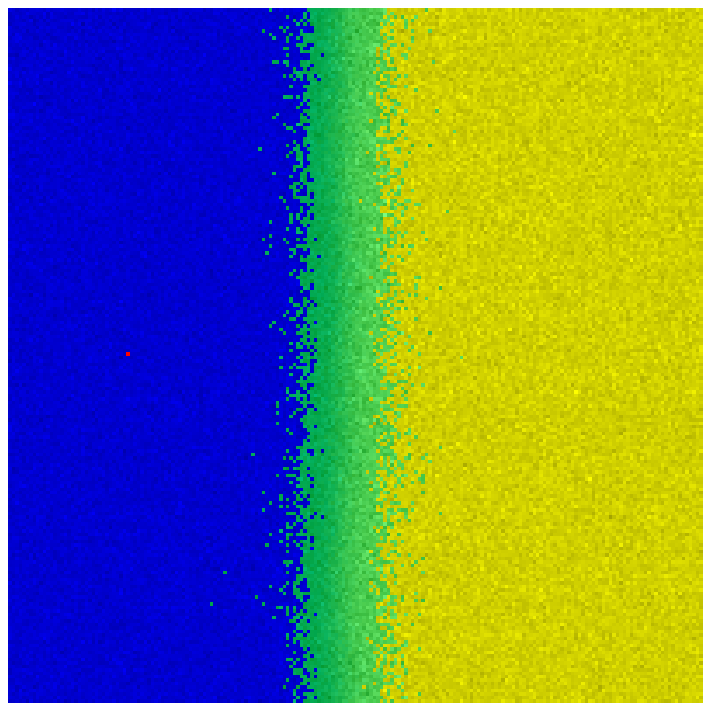,
        width=0.4\linewidth,clip=,angle=0}
}} \vspace{.1cm} \centerline{\mbox{ \hfill
\epsfig{file=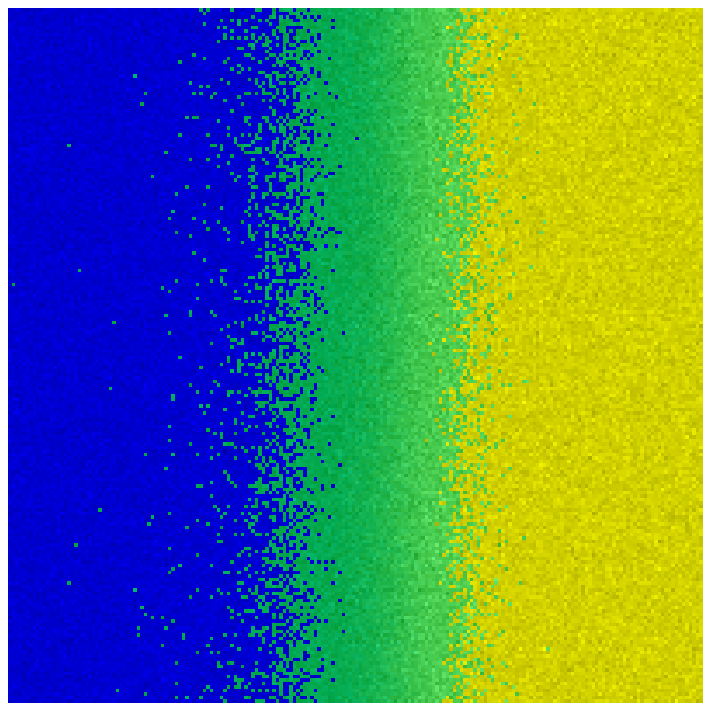,
        width=0.4\linewidth,clip=,angle=0}
\hspace{.05cm} \epsfig{file=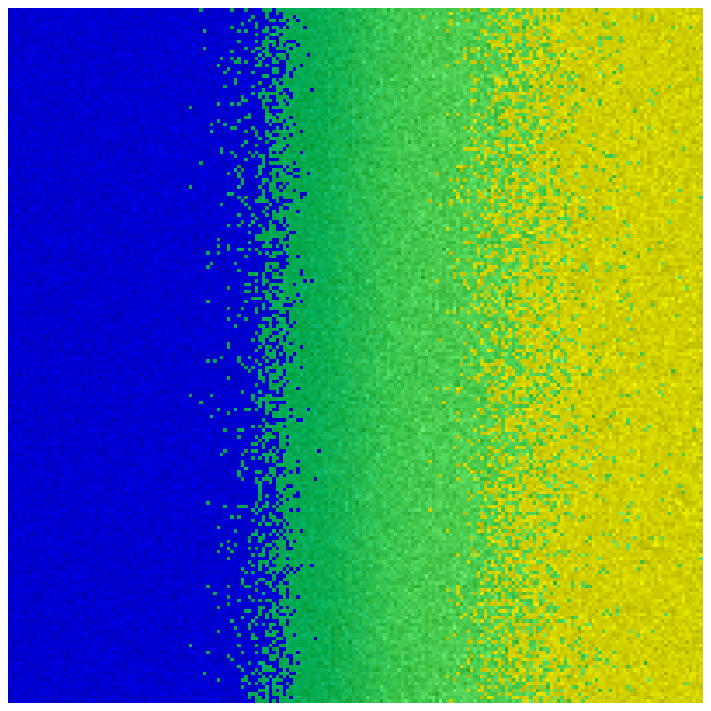,
        width=0.4\linewidth,clip=,angle=0}
}} \caption{Concentration field at a given time instant for
$k=0.0005$ (top left panel) and $k=0.001$ (top right panel). The
system size is $200 \times 200\times5$. Lower panels show the
concentration field for $k=0.0005$ and $\gamma_A=0.25$,
$\gamma_B=1$ (left), and $k=0.0005$, $\gamma_A=1$, $\gamma_B=0.25$
(right). The structure of the reaction zone can be seen in these
figures.
 \label{frontpic1}}
\end{figure}

Starting from this initial condition a reaction front will develop
as the autocatalyst $B$ consumes the fuel $A$ in the reaction. The
front will move with velocity $c$ and it is convenient to study
the front dynamics in a frame moving with velocity $c$.
Propagating fronts are depicted in Fig.~\ref{frontpic1}, which
shows the concentration field at a given time instant. The upper
two panels plot the front for two values of the reaction rate
constant $k$ and $\gamma_A=\gamma_B=1$.
We see that for $k=0.0005$ the front profile is much thicker than
that for $k=0.001$. This dependence is in accord with predictions
based on a reaction-diffusion description of the front as can be
seen from the analysis given below.

\begin{figure}[!htbp]
\centerline{\mbox{
\epsfig{file=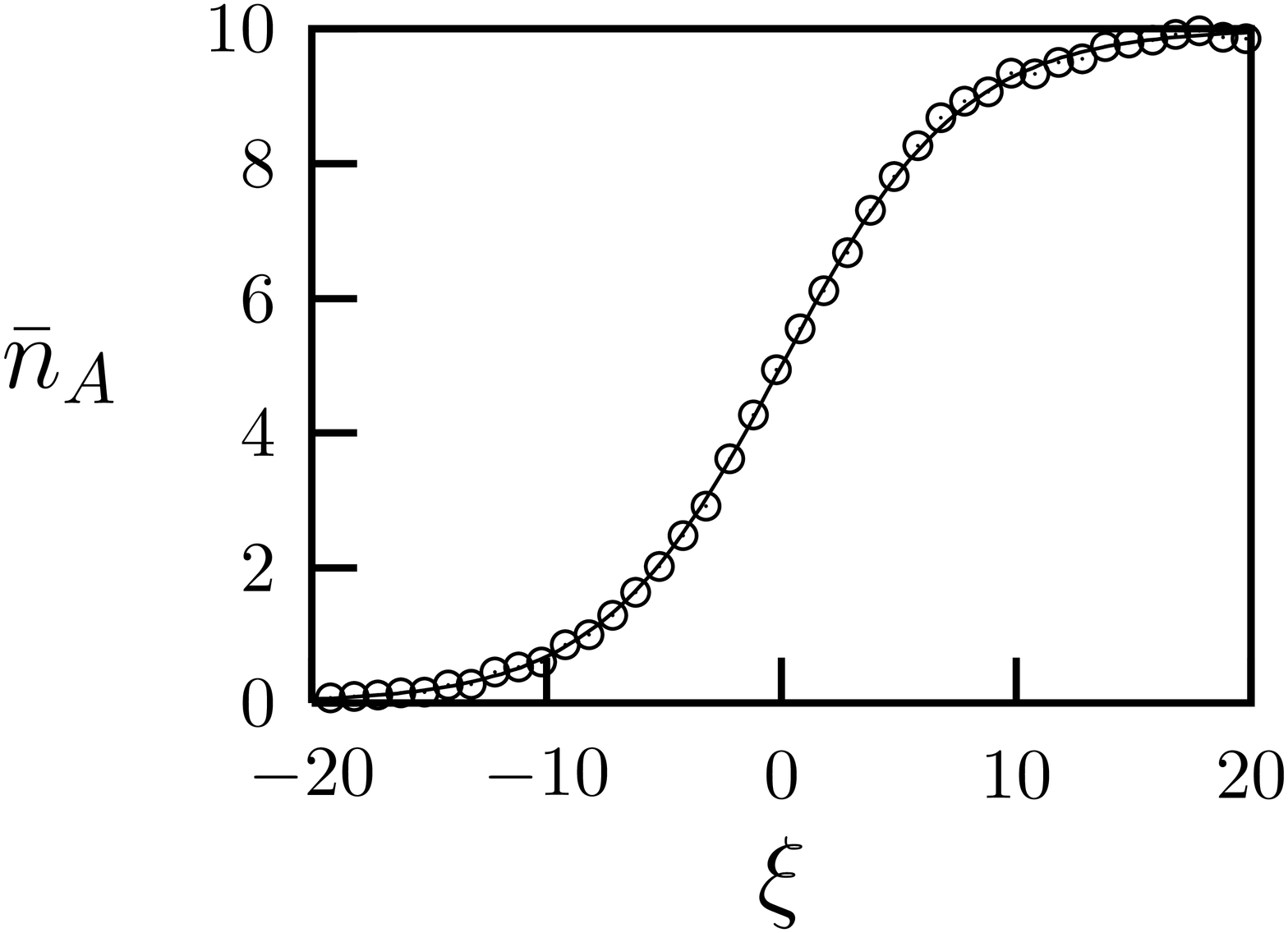,
        width=0.40\linewidth,clip=,angle=90}
\hspace{-5mm}
\epsfig{file=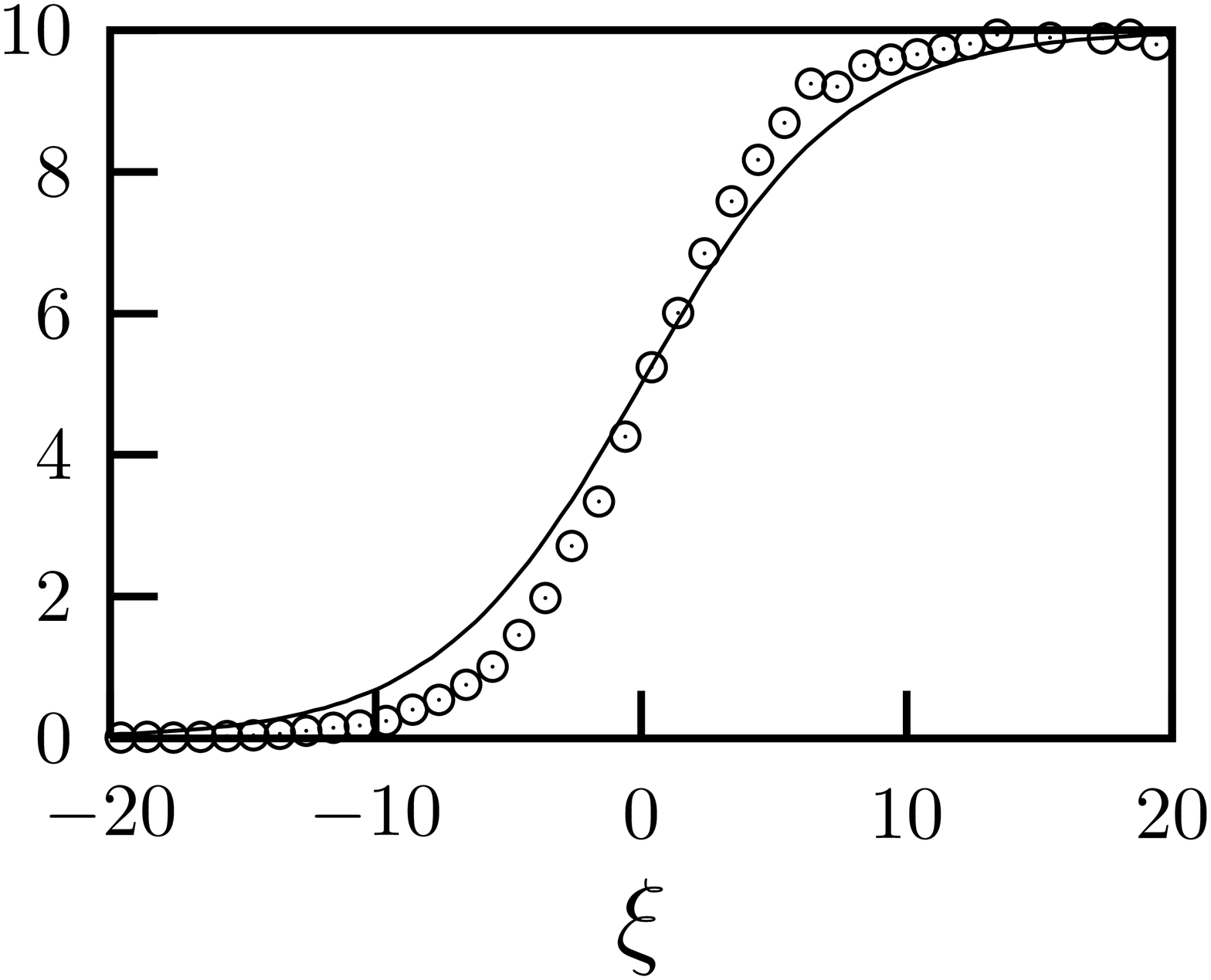,
        width=0.40\linewidth,clip=,angle=90}
}} \caption{Plot of the front profile ${\bar n_A}(\xi)$ versus
$\xi$ for different values of $k$: left panel, $k=0.0005$; right
panel, $k=0.001$. The system size is $200\times200\times5$.  The
continuous line represents the theoretical value obtained from
Eq.~(\ref{eq:profile}). \label{front}}
\end{figure}

The structure of these planar fronts can investigated
quantitatively by studying the front dynamics in a frame moving
with the front velocity, $\xi=x-ct$, and averaging the
concentration profile over the width (along y) of the front,
$\bar{n}_A(\xi)=\int dy\; n_A(\xi,y)$. Figure~\ref{front} plots
$\bar{n}_A(\xi)$ for the two values of $k$ used in
Fig.~\ref{frontpic1}.
>From this figure we see that a well-defined propagating reaction
front is obtained and the width of the front decreases as the
reaction rate increases relative to the diffusion rate.

The front shape and velocity can be determined from the reaction-diffusion
equation. For a planar front propagating along the $x$-direction, in a frame
moving with the front velocity, the reaction-diffusion
equation~(\ref{eq:rd}) is,
\begin{eqnarray}
D \frac{d^2 }{d \xi^2} \bar{n}_A(\xi)+c\frac{d }{d
\xi}\bar{n}_A(\xi) -k \bar{n}_A(\xi) (\bar{n}_0-\bar{n}_A(\xi))^2 =0 \;.
\nonumber \\
\end{eqnarray}
The front profile can be found analytically from the solution
of this equation and is given by
\cite{mck}
\begin{equation}\label{eq:profile}
\bar{n}_A(\xi)= \bar{n}_0 \Big( 1+ e^{-c \xi/D}\Big)^{-1}\;,
\end{equation}
where the front speed is $c=(Dk\bar{n}_0^2/2)^{1/2}$. The profile
for species $B$ can be found from the conservation condition
$\bar{n}_A+\bar{n}_B=\bar{n}_0$. Figure~\ref{front} compares this
analytical prediction with the simulation results of the MPC
reaction-diffusion dynamics. For $\gamma_A=\gamma_B=1$ the mutual
diffusion coefficient $D$ is given by
Eq.~(\ref{eq:diffusionCoeficient}). There is good agreement
between the simulation and analytical values for small $k$ where
the conditions for the validity of the mean field approximation
are satisfied. For larger $k$ values, such as $k=0.001$ in the
right panel of the figure we see that there are deviations from
the mean field result. For this value of $k$, the reaction is fast
and there is a breakdown of the local Poissonian equilibrium and a
reaction-diffusion description is not applicable. A similar
breakdown is observed for very small $k$, for example for
$k=0.0002$, due to the fact that very few reactive events occur in
the reaction zone of the front and fluctuations are important.

\begin{figure}[htbp]
\centerline{\mbox{
\epsfig{file=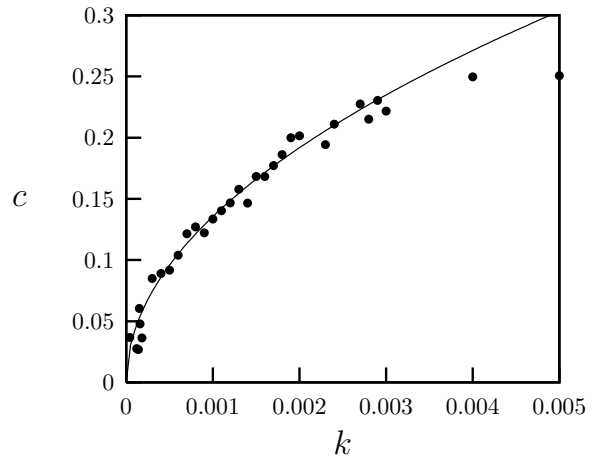,width=0.75\linewidth,clip=,angle=90} }}
\caption{ (filled circles) Front velocity $c$ as a function of
$k$. (solid line) Front velocity from Eq.~(\ref{eq:profile}).
  \label{fig:c-vs-k} }
\end{figure}

The front velocity was determined from the simulation data as a
function of $k$.  In Fig.~\ref{fig:c-vs-k} we plot the front
velocity $c$ versus $k$ and compare the simulation results with
the prediction $c=(Dk\bar{n}_0^2/2)^{1/2}$.
The front velocity agrees with the simulation results for $k \le
0.003$, although the front profile deviates slightly from the
predicted value for somewhat smaller values of $k$ ($ k \le
0.001$).

More microscopic aspects of the front structure and dynamics that
are captured by the MPC model are illustrated in the lower two
panels of Fig.~\ref{frontpic1}. These figures plot snapshots of
the front for $k=0.0005$, the same value of $k$ as in the top left
panel of the figure, but for two different pairs of
$\gamma_\alpha$ values, $(\gamma_A=0.25, \gamma_B=1.0)$ and
$(\gamma_A=1.0, \gamma_B=0.25)$. Comparison of the lower panels of
the figures, and also with the upper left panel, shows that the
structures of the interfacial zones are different.  In the MPC
dynamics employed here, the diffusion of the species depends on
their density and $\gamma_\alpha$. Since the density of the
species changes significantly in the interfacial zone, it is
likely that a concentration-dependent mutual diffusion coefficient
is required to describe this structure.

\section{Conclusion} \label{sec:conc}

The generalizations of the multi-particle collision model
described here, and its extensions, allow one to study a variety
of phenomena at the mesoscopic level. In particular, the ability
to simulate the dynamics of multi-component systems whose
diffusion coefficients can be different means that
diffusion-driven instabilities, such as the transverse cubic
autocatalytic front instability considered in this paper, can be
investigated. Since the mesoscopic MPC model preserves the basic
conservation laws in the system, to study such instabilities
requires the presence of a third solvent species so that there are
two independent diffusion coefficients in the system. The method
could also be used to study reactive and non-reactive binary fluid
{\em flows} which also show interesting instabilities where
fluctuations play a role near the onset of instabilities.

The cubic autocatalytic reaction is simply one example of a much
broader class of reaction-diffusion systems that can be studied
using reactive versions of the mesoscopic multi-particle collision
dynamics. In particular, more general reaction-diffusion dynamics
in specific geometries relevant for the materials science and
biological applications may be carried out. The
presence of flows can also be treated easily in this context.

While we have focused primarily on parameter domains where mean
field approximations are largely applicable, one of the most
interesting applications of the methodology introduced in this
paper is to systems on mesoscales where particle numbers are small
so that fluctuations play a crucial role in the dynamics and
system geometry is important.

Acknowledgements: This work was supported in part by a grant from
the Natural Sciences and Engineering Research Council of Canada
and in part by the grant C-1279-0402-B from Consejo de Desarrollo 
Científico Humanístico y Tecnológico of Universidad de Los Andes.


\begin{thebibliography}{99}

\bibitem{nicolis} G. Nicolis and I. Prigogine, {\em
Self-Organization in Non-Equilibrium Systems}, (Wiley, New York,
1977).

\bibitem{mpc1} A. Malevanets,
R. Kapral, Europhys. Lett., \textbf{44}(5), 552 (1998).

\bibitem{mpc2} A. Malevanets and R. Kapral, J. Chem. Phys., {\bf
110}, 8605 (1999).

\bibitem{mpc3} A. Malevanets and R. Kapral, J. Chem. Phys., {\bf
112}, 7260 (2000).

\bibitem{mesofin} A. Malevanets and R. Kapral, Lect. Notes Phys. {\bf 640},
113 (2004).

\bibitem{flows} T. Ihle, D. M. Kroll, Phys. Rev. E \textbf{63}, 020201
(2001); A. Lamura, G. Gompper, T. Ihle, D. M. Kroll, Europhys.
Lett., \textbf{56}, 768 (2001); A. Lamura, G. Gompper, T. Ihle, D.
M. Kroll: Europhys. Lett., \textbf{56}, 319 (2001).

\bibitem{colloids} Y. Hashimoto, Y. Chen, H. Ohashi, Comp. Phys. Comm.,
\textbf{129}, 56 (2000); Y. Inoue, Y. Chen, H. Ohashi, Colloids
and Surfaces A, \textbf{201}, 297 (2002); T. Sakai, Y. Chen, H.
Ohashi, Phys. Rev. E, \textbf{65}, 031503 (2002).

\bibitem{polymers} A. Malevanets, J. M. Yeomans, Europhys. Lett.,
\textbf{52}, 231 (2000).

\bibitem{songhi} S. H. Lee and R. Kapral, J. Chem. Phys., {\bf
121}, 11163 (2004).

\bibitem{kay} K. Tucci and R. Kapral, J. Chem. Phys., {\bf 120},
8262 (2004).

\bibitem{yeomans} A. Malevanets and J. M. Yeomans, Comp. Phys. Commun., {\bf 129},
282 (2000).

\bibitem{gardiner} C. W. Gardiner, {\em Handbok of Stochastic
Processes}, (Springer-Verlag, New York, 1985).

\bibitem{ebeling} R. Kapral, in {\em Stochastic Dynamics}, eds.,
L. Shimansky-Geier and T. P\'oschel, (Springer, Berlin, 1997), p.
294.

\bibitem{cub1} D. Horv\'ath and K. Showalter, J. Chem. Phys.
{\bf 102}, 2471 (1995).

\bibitem{cub2}
J. Billingham and D. J. Needham, {\em Phil. Trans. R. Soc., Ser.
A} {\bf 334}, 1 (1991).

\bibitem{cub3} S. K. Scott and K. Showalter, J. Phys. Chem. {\bf 96},
8702 (1992).

\bibitem{cub4}
D. Horv\'ath, V. Petrov, S. K. Scott and K. Showalter, {\em J.
Chem. Phys.} {\bf 98}, 6332 (1993).

\bibitem{cub5} Z. Zhang and {S. A. E. G.} Falle, Proc.R. Soc., Ser. A
{\bf 446}, 1 (1994).

\bibitem{cub6} {R. A.} Milton and {S. K.} Scott, J. Chem. Phys.
{\bf 102}, 5271 (1995).

\bibitem{mck} A. Malevanets, A. Careta and R. Kapral, Phys. Rev. E, {\bf 52},
4724 (1995).

\bibitem{deGM} S. R. de Groot and P. Mazur, {\em Nonequilibrium
Thermodynamcis}, (North-Holland, Amsterdam, 1962).


\end{thebibliography}
\end{document}